\documentclass[aps,prl,twocolumn,showpacs,amsmath,amssymb]{revtex4}
\usepackage{graphics}
\usepackage{epsfig}
\def\be{\begin{equation}}
\def\ee{\end{equation}}
\def\la{\langle}
\def\ra{\rangle}
\begin{document}
\title{Continuous quantum measurement with independent
detector cross-correlations}
\author{Andrew N. Jordan and  Markus B\"uttiker}
\affiliation{D\'epartement de Physique Th\'eorique, Universit\'e de Gen\`eve,
CH-1211 Gen\`eve 4, Switzerland} \date{May 2, 2005}
\begin{abstract}
We investigate the advantages of using two independent, linear
detectors for continuous quantum measurement.  For single-shot quantum
measurement, the measurement is maximally efficient if the detectors
are twins.  For weak continuous measurement, cross-correlations allow
a violation of the Korotkov-Averin bound for the detector's
signal-to-noise ratio.  A vanishing noise background provides a
nontrivial test of ideal independent quantum detectors.  We further
investigate the correlations of non-commuting operators, and consider
possible deviations from the independent detector model for mesoscopic
conductors coupled by the screened Coulomb interaction.
\end{abstract}

\maketitle
There has recently been intensive research,
both experimental and theoretical, into the development
of quantum detectors in the solid state.
Mesoscopic structures, such as the quantum point
contact, single electron transistor, and SQUID have been
used for fast qubit read-out.
Contrary to the historical assumption that
the quantum measurement occurs instantaneously,
in the modern theory of quantum detectors the continuous
nature of the measurement process is essential to the understanding
and optimization of how quantum information is collected.

The ultimate goal for quantum computation is the development of
``single-shot'' detectors, where in one run the qubit's state is
unambiguously determined.  An important figure of merit is the
detector's efficiency, defined as the product of the time taken to
measure the state of the qubit, with the measurement induced dephasing
rate.  This efficiency is minimized for detectors that do not lose
information about the quantum state as the measurement is being
performed.  Single-shot detectors are difficult to realize because of
the fast time resolution required.  Another approach is that of weak
measurement, where detector backaction renders the state of the qubit
invisible in the average output of the detector, but quantum coherent
oscillations are uncovered in the spectral density of the detector.
These measurements are easier to preform because both detector and
qubit averaging are permitted, and the experiments only require a
bandwidth resolution of the qubit energy splitting.  One important
result in weak measurement theory is the Korotkov-Averin bound.  It
states that as the weak measurement is taking place, the detector
backaction quickly destroys the quantum oscillations, so the maximum
detector signal-to-noise ratio is fundamentally limited at 4.  This
result was derived in Refs.~\cite{stn}, confirmed in
Refs.~\cite{verifystn}, generalized in Refs.~\cite{generalized}, and
measured in Ref.~\cite{expt}.

In this Letter, the theoretical advantages of considering the
cross-correlated output of independent quantum detectors are
investigated.  It is clear that cross-correlations bestow an
experimental advantage \cite{beuhler} in quantum measurement, because
the procedure filters out any noise not shared by the two detectors.
Thus, extraneous noise produced by sources such as charge traps in one
detector will be removed.  This technique is also used in quantum
noise measurements for this same advantage \cite{noise}.  We demonstrate that
although the two detectors cannot improve the efficency of the
detection process, the cross-correlated output can violate the
Korotkov-Averin bound.  As a solid-state implementation of the results
derived, Fig.~1 depicts two quantum point contacts capacitively
coupled to the same double quantum-dot representing a charge qubit.
It should be stressed that such structures have already been
fabricated \cite{DD}.

\begin{figure}[b]
\begin{center}
\leavevmode
\psfig{file=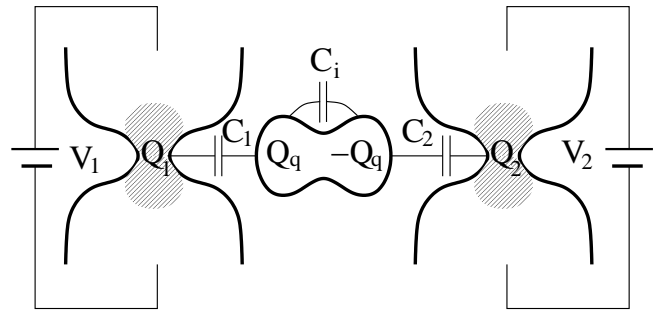,width=8.6cm}
\caption{Cross-correlated quantum measurement set-up:
Two quantum point contacts are measuring the same double quantum dot
qubit.  As the quantum measurement is taking place, the current
outputs of both detectors can be averaged or cross-correlated with each other.}
\label{geometry}
\end{center}
\vspace{-5mm}
\end{figure}

{\it Detector assumptions and linear response.}-- 
We employ the linear response approach to quantum measurement 
because of its elegant simplicity, and general
applicability to a wide range of detectors \cite{stn,pilgram,stone}. 
The quantum operator to be
measured is $\sigma_z$.  The Hamiltonian is
\begin{equation}
H = -(\epsilon\, \sigma_z +\Delta \sigma_x)/2 +
H_1 + H_2 + Q_1 \sigma_z/2 + Q_2 \sigma_z/2,
\label{ham}
\end{equation}
where $Q_{1,2}$ are the bare input variables of detector $1$ and $2$, and
$H_{1,2}$ are their Hamiltonians, and $I_{1,2}$ are the bare output
variables of the detectors.  The small coupling constants are
incorporated into the definition of $Q_{1,2}$.  We assume that the
detector is much faster than all qubit time scales, so the relevant
detector correlation functions are the stationary zero frequency
correlators:
\begin{subequations}
\begin{eqnarray}
&&\langle I_{i}(t + \tau ) I_{j}(t) \rangle \;= \;
S_{I}^{(i)}\delta_{ij} \delta (\tau),
\label{SI}\\
&&\langle Q_{i}(t + \tau ) Q_{j}(t) \rangle =
S_{Q}^{(i)}\delta_{ij} \delta (\tau),
\label{SQ}\\
&&\langle Q_{i}(t + \tau ) I_{j}(t) \rangle =
({\rm Re}S^{(i)}_{Q I} + i\, {\rm Im}S^{(i)}_{Q I})
\delta_{ij} \delta (\tau),
\label{SQI}\\
&&\langle I_{i}(t + \tau ) Q_{j}(t) \rangle =
({\rm Re}S^{(i)}_{Q I}
- i\, {\rm Im}S^{(i)}_{Q I})\delta_{ij} \delta (\tau).
\label{SIQ}
\end{eqnarray}
\end{subequations}
where the time delta functions have a small shift $\delta (\tau-0)$,
reflecting the finite response time of the detector.
Linear response theory tells us that the response coefficients
$\lambda_{1,2}$ are given by
$\lambda_{i} = -2\, {\rm Im}S^{(i)}_{Q I} /\hbar$,
 so the output of the detectors (with the average subtracted) is
${\cal O}_{i} = I_{i} +  \lambda_{i} \sigma_z/2$.
As the detector is turned on, it ideally collects information about the
operator $\sigma_z$ while destroying information about $\sigma_{x,y}$.
The measurement is complete when the integrated difference in qubit
signal exceeds the detector noise, so the state of the qubit may be
determined.  In the simplest case of $\Delta=0$, the standard
expressions for the dephasing rate $\Gamma$ and measurement time $T_M$
are \cite{rmp}
\be
\Gamma = S_Q/(2\hbar^2), \quad T_M =4 S_I/ \lambda^2.
\label{relations}
\ee
Let us next observe
\begin{equation}
\hbar^2 \lambda^{2}= 4({\rm Im} S_{QI})^2 \le 4 \vert S_{QI} \vert^2 \le
4 S_Q S_I,
\label{lr}
\end{equation}
where we have used the Cauchy-Schwartz inequality.
For a lone detector the above relations imply
$\Gamma T_M \ge 1/2$,
where equality is reached for quantum limited detectors.
The two conditions needed to reach the ``Heisenberg efficiency'' are,
\be
(a)\;\; {\rm Re} S_{QI} = 0, \qquad
(b)\;\; \vert S_{QI} \vert^2 = S_Q S_I.
\label{he}
\ee In the context of mesoscopic scattering detectors, condition (b)
is related to the energy dependence of the transmission of the
scatterer, while (a) is related to the symmetry of the scatterer \cite{stn}.
Pilgram and one of the authors derived Eqs.~(\ref{he}) for arbitrary
detectors described by scattering matrices \cite{pilgram}.  Clerk, Girvin, and
Stone interpreted these conditions as no lost information either
through (a) phase or (b) energy averaging \cite{stone}.

{\it Can we do better with two detectors?}-- By adding an additional
detector to the qubit, the signals may be averaged, ${\cal O}=({\cal
O}_1+{\cal O}_2)/2$, so the measurement time may be reduced.  On the
other hand, the new detector dephases the qubit more quickly.  For
statistically independent detectors, the measurement-induced dephasing
rate is simply the sum of the individual dephasing rates, so the
two-detector efficiency is
\be \Gamma T_M = 2
(S_I^{(1)}+S_I^{(2)})(S_Q^{(1)}+S_Q^{(2)})/\hbar^2
(\lambda_1+\lambda_2)^2 \ge 1/2,
\label{averaged}
\ee where equality is reached for twin detectors that are themselves
quantum limited.  This condition may also be interpreted as no lost
information between the two detectors.  Rather than average the
signals, we could instead cross-correlate them.  However, this also brings
no advantage because the new signal obtained by multiplying the output
from the two detectors, ${\cal O}_{1} (t_1) {\cal O}_{2}(t_2 )$, has
its own noise.  If we could average over many trials the noise could
be eliminated, but for single-shot measurement the efficiency is still
intrinsically limited.

{\it Violation of the Korotkov-Averin bound.}-- But consider next
Korotkov and Averin's bound on the signal-to-noise ratio for a weakly
measured qubit \cite{stn}.  It states that the ratio of the measured
qubit signal to detector noise, $\cal R$, is fundamentally limited by
4.  This bound can be overcome with quantum nondemolition measurements
by increasing the signal \cite{qnd}.  In this Letter, we are concerned
with reducing the noise.  To see how this bound emerges, we briefly
derive this inequality for one detector.  The Hamiltonian is given by
Eq.~(\ref{ham}) with $Q_2=0$.  The time averaged output of the
detector is $\la{\cal O}\ra= (\lambda/2) (1/T) \int_{0}^{T} dt \langle
\sigma_z (t) \rangle$.  For a weakly measured qubit, the statistical
average over $\sigma_z$ is taken with respect to the stationary,
mixed, density matrix of the qubit, $\rho=(1/2)\openone$,
and therefore the qubit makes {\it no} contribution to the average
output current.  The detector's spectral density is $S(\omega) = S_{I}
+ (\lambda^{2}/4) S_{zz}(\omega)$, where \be S_{ij}(\omega) = 2
\int_{0}^{\infty} dt \cos(\omega t) \langle \sigma_i (0) \sigma_j (t)
\rangle.
\label{Sij}
\ee The qubit dynamics may be found by expanding the evolution
operator to second order in the coupling constant, and averaging over
the $\delta$-correlated $Q$ fluctuations to obtain equations of
motion, with dephasing rate $\Gamma$.  For the special case of
$\epsilon=0$, the noise spectrum in the vicinity of
$\omega=\Delta/\hbar \equiv \Omega$ is \cite{stn}
\begin{equation}
S(\omega) = S_{I} + \frac{\lambda^2 \Gamma}{2} \frac{
 \Omega^2}{(\omega^2-\Omega^2)^2 + \omega^2 \Gamma^2}.
\end{equation}
At the qubit frequency, $\omega = \Omega$, the signal has a maximum of
$S_{\rm max} = \lambda^2/(2 \Gamma)= \hbar^2\lambda^2/S_{Q}$.  We use
again the linear response relation (\ref{lr}) to bound the
signal-to-noise ratio of the detector as 
\be 
{\cal R} = S_{\rm max}/S_I \le 4.
\label{STN}
\ee
This is the Korotkov-Averin bound.

Consider now the cross-correlation of the outputs from two independent
detectors, both measuring the same qubit operator $\sigma_z$.  The
qubit dynamics is the same, except that $\Gamma=\Gamma_1+\Gamma_2$.
The spectral density of the cross-correlation $S_{1,2}(\omega)$
contains four terms,
\begin{eqnarray}
S_{1,2}(\omega)&=&\int_{0}^{\infty}dt\cos(\omega t)
[2 \langle I_{1} (0) I_{2} (t) \rangle 
+  \lambda_1 \langle\sigma_z (0)  I_2(t)  \ra  \nonumber \\
&+& \lambda_2\la  I_1(0) \sigma_z (t) \rangle 
+(\lambda_1\lambda_2/2) \langle  \sigma_z (0) \sigma_z (t) \rangle ].
\label{s12}
\end{eqnarray}
According to Eq.~(\ref{SI}) the bare detector noise of the two
detectors are uncorrelated, the qubit dynamics is uncorrelated with
the bare detector noise, so only the qubit signal (\ref{Sij})
contributes to the correlation function (\ref{s12}).  The remaining
question is the detector configuration that maximizes the signal.  The
maximum signal at $\omega =\Omega$ is $S_{\rm max}=\lambda_1\lambda_2
/[2 (\Gamma_1+\Gamma_2)]$, and we may use the relations (\ref{lr}) to
bound the cross-correlated signal in relation to the noise of the
individual detectors as \be S_{\rm max}
\le  2 \sqrt{S_I^{(1)}S_I^{(2)} },
\label{maxS}
\ee where equality is reached for $S_Q^{(1)}=S_Q^{(2)}$.  For twin
detectors, (\ref{maxS}) is half of the single detector signal, because
of the doubled dephasing rate \cite{note1}.  

We have successfully removed the background noise, and can now see the
naked destruction of the qubit.  The signal-to-noise ratio ${\cal R}$
is divergent, violating the Korotkov-Averin bound.  Why did
cross-correlations help here, but not in the quantum efficency?  The
reason is that the spectral density, in contrast to the measurement
efficiency, is not protected by the uncertainty principle, so there is
no fundamental limitation on its measurement. 

{\it Detecting the detector.}--- Although the above result is very
appealing, one might worry that it can be spoiled by some weak direct
coupling between the detectors.  We now take this into account by
introducing another term in the Hamiltonian, $H_{1,2} = \alpha Q_1
Q_2$, where $\alpha$ is a relative coupling constant between the two
detectors.  The additional contribution to the zero-frequency
cross-correlator, $\delta S_{1,2} = \int dt \langle \delta {\cal
O}_{1}(0) \delta {\cal O}_{2}(t) \rangle$, consists of four terms,
\begin{eqnarray}
&&\delta S_{1,2}=\int dt [\langle I_1(0) I_2(t) \rangle
+ \alpha \lambda_1 \langle
 Q_2(0) I_2(t) \rangle  \nonumber  \\
&& + \alpha \lambda_2 \langle I_1(0) Q_1(t) \rangle
+ \alpha^2 \lambda_1 \lambda _2 \langle Q_1(0) Q_2(t) \rangle].
\end{eqnarray}
The first and last term vanish for independent detectors,
leaving the middle two terms.  These middle terms may be interpreted
as a fluctuation in one detector causing a response in the
other detector, which is then correlated back with the
original bare detector variable in the signal multiplication.
Using the correlators Eq.~(\ref{SQI},\ref{SIQ}), we find
\be
\delta S_{1,2} =
\alpha \lambda_1 {\rm Re} S^{(2)}_{QI}
   + \alpha \lambda_2 {\rm Re} S^{(1)}_{QI},
\label{addcross}
\ee where we have substituted the response coefficient for the
imaginary part of the $Q$-$I$ correlator, which causes the imaginary
part of the additional cross-correlated signal to vanish.  An
interesting aspect of the above equation is that if the detectors are
both quantum limited, we saw in Eq.~(\ref{he}a) that one condition was
that the real part of the $Q$-$I$ correlator should vanish.
Eq.~(\ref{addcross}) provides a simple experimental test to check this
condition: background cross-correlations should vanish.  If this is
true, then the direct coupling between detectors does not give any
noise pedestal to overcome for weak measurement.
\begin{figure}[t]
\begin{center}
\leavevmode
\psfig{file=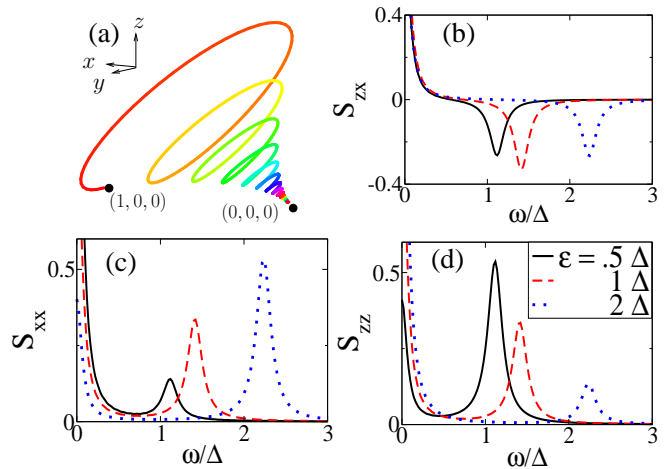,width=8.6cm}
\caption{(color online). (a) Time domain destruction of the quantum
state by the weak measurement process for $\epsilon =\Delta$.  The
elapsed time is parameterized by color, and (x,y,z) denote coordinates
on the Bloch sphere.  (b) The measured cross-correlator
$S_{zx}(\omega)$ changes sign from positive at low frequency
(describing incoherent relaxation) to negative at the qubit
oscillation frequency (describing out of phase, coherent
oscillations).  (c,d) The correlators $S_{xx}$, $S_{zz}$ have both a
peak at zero frequency and at qubit oscillation frequency.  We take
$\Gamma=\Gamma_x=\Gamma_z= .07 \Delta/\hbar$.  $S_{ij}$ are plotted in
units of $\Gamma^{-1}$.}
\label{combo}
\end{center}
\vspace{-5mm}
\end{figure}

{\it Weak measurement of non-commuting observables}.--- Once we have
two detectors involved, there is no reason why they both have to
measure the same observable (or one that commutes with it).  We now
consider an experiment where one detector measures $\sigma_z$ and the
other measures $\sigma_x$, and the outputs are cross-correlated.  The
measured spectrum is $S_{1,2}(\omega) = (\lambda_1 \lambda_2/4)
S_{zx}(\omega)$.  This experiment could be implemented with a split
Cooper-pair box \cite{CPB}, where a SQUID is weakly measuring the
persistent current, and a quantum point contact is weakly measuring
the electrical charge.  In standard measurement theory, the question of a
simultaneous measurement of non-commuting observables cannot even be posed.

The coupling part of the Hamiltonian is altered
to be $H_c= (1/2) Q_1 \sigma_z + (1/2) Q_2 \sigma_x$.  
We parameterize any traceless qubit operator as $\sigma(t)
=\sum_i x_i(t) \sigma_{x_i}$, and the density matrix $\rho =
(1/2)\openone + \sigma(t)$, so $(x, y, z)$ also represent coordinates
on the Block sphere.  Defining $\Gamma_z = S_Q^{(2)}/(2 \hbar^2)$,
$\Gamma_x = S_Q^{(1)}/(2 \hbar^2)$, the equations of motion for $x_i$,
averaged over the white noise of $Q_1$ and $Q_2$, are
\be
\begin{pmatrix}
\dot x \\ \dot y \\ \dot z
\end{pmatrix}
= \begin{pmatrix}
-\Gamma_z & -\epsilon/\hbar & 0 \\
 \epsilon/\hbar & -\Gamma_x -\Gamma_z & -\Delta/\hbar \\
0 & \Delta/\hbar & -\Gamma_x
\end{pmatrix}
\begin{pmatrix}
x \\  y \\  z
\end{pmatrix}.
\label{block-red}
\ee 
Diagonalization of the transition matrix in the case $\Gamma_x
=0$ gives the usual expressions for the dephasing and relaxation
rates.  This set-up is always far away from the Heisenberg
efficiency because one detector is destroying the signal
the other is trying to measure.  However, this situation
has interesting behavior in the weak measurement case.
The cross-correlation $S_{1,2}(\Omega)$ attains
its maximum signal at the symmetric point $\epsilon = \Delta$,
$\Gamma_x=\Gamma_z=\Gamma$, so the qubit frequency is
$\Omega=\sqrt{2}\Delta/\hbar$.  The master equation may be solved in the
weak dephasing limit, giving the correlation (for positive
frequencies)
\be S_{xz}(\omega)  = S_{zx}(\omega) = \frac{\Gamma}{\Gamma^2 + \omega^2}
- \frac{3\Gamma} {9\Gamma^2 + 4 (\omega -\Omega)^2}.
\label{Szx}
\ee The first term has a peak at zero frequency, while the second term
has a peak at $\omega =\Omega$, with width $3\Gamma/2$, and signal
$-1/3 \Gamma$.  Bounding this signal in relation to the noise in the
individual twin detectors gives $\vert S_{1,2}(\Omega) \vert \le (2/3) S_I$.
The interesting feature of this correlator is that it changes sign as
a function of frequency.  The low frequency part describes the
incoherent relaxation to the stationary state, while the high
frequency part describes the out of phase, coherent oscillations of
the $z$ and $x$ degrees of freedom.  The measured correlator
$S_{zx}$, as well as $S_{xx}, S_{zz}$ are plotted as a function of frequency
in Fig.\ref{combo}(b,c,d) for different values of $\epsilon$.  These
correlators all describe different aspects of the time domain
destruction of the quantum state by the weak measurement, visualized
in Fig.\ref{combo}a.  We note that the cross-correlator changes
sign for  $\epsilon = -\Delta$.

{\it Implementation.}-- We now consider two quantum point contacts,
measuring a double quantum dot qubit.  The point contact perfectly
obeys conditions (\ref{he}) and is thus an ideal detector
\cite{stn,pilgram,stone}.  The bare input detector variable $Q$ is
identified with the electrical charge in the point contact, while
the bare output variable $I$ is identified with the shot noise.  The
conductance of the QPC is sensitive to the electron's position on the
double dot.  A measurement of the quantum state occurs when the
integrated current difference exceeds the shot noise power.  In the
geometry shown in Fig.~1, one detector measures $\sigma_z$, while
the other detector measures $-\sigma_z$, so the qubit signal
will be anti-correlated.  The charges on the two detectors are not
independent, but rather must be the opposite of each other to have
charge neutrality in the system.  This electrical screening generates
correlations between the potentials of the two quantum point contacts,
increasing the dephasing rate, hurting the efficiency, and
producing some background noise for the cross-correlator.  This
situation is markedly in contract with the single detector case
\cite{pilgram}, where screening simply renormalized the coupling
constant.  However, in realistic detectors there will always be other
gates to control the quantum double dot, creating a larger capacitance
matrix than the minimal one shown in Fig.~1.  In this extended
geometry, a charge fluctuation in one detector will be screened by the
surrounding metallic gates, not by the other detector, justifying the
independent detector model.  We mention that in the experiment already
done by Buehler {\it et al.} \cite{beuhler}, the detectors seem to be
completely independent.

{\it Conclusions.}-- We considered the advantages that two independent
quantum detectors measuring the same qubit can bring to the quantum
measurement problem.  The Heisenberg efficiency could be
reached with quantum limited twin detectors.  The asymmetry of the
detector, related to phase information in the case of mesoscopic
scattering detectors, could be measured with low-frequency
cross-correlations, and thus provides a non-trivial experimental test
for quantum efficiency.  For weak continuous measurement, the
cross-correlated signal removes the noise pedestal, and allows a
violation of the Korotkov-Averin bound on the signal-to-noise ratio.
The cross-correlation of non-commuting operators were also
investigated, which showed a cross-over from positive to negative
correlation as a function of frequency.  Although we have focused on
mesoscopic qubits, this technique easily extends to other systems
where similar bounds have been derived, such as single spins and
nano-mechanical oscillators \cite{more}.

We thank A. N. Korotkov and D. V. Averin for correspondence and suggestions. 
This work was supported by the SNF and MaNEP.


\begin{thebibliography}{99}
\bibitem{note1}
The response functions may be quite different,
so the cross-correlated signal $S_{\rm max} \le
2 S_I^{(2)} \lambda_1/\lambda_2 = 2 S_I^{(1)} \lambda_2/\lambda_1$
may be much larger than the noise in one detector,
provided it is much smaller than the noise in the other detector.
\end{thebibliography}
\end{document}